\documentclass[12pt]{article}
\usepackage{amsmath}
\usepackage{amsfonts}
\usepackage{amssymb}
\usepackage[left=2cm,right=2cm,
top=2cm,bottom=2cm,bindingoffset=0cm]{geometry} %

\begin{document}
\title{Coupled Cavity Model:  Correctness and Limitations} 
\date{}
\author{M.I.Ayzatsky\thanks{mykola.aizatsky@gmail.com}\\
National Science Center\\Kharkov Institute of Physics and Technology (NSC KIPT),\\610108, Kharkov, Ukraine}
 
 \maketitle
 
\begin{abstract}
Results of analysis of correctness and limitations of the classical Coupled Cavity Model are presented in the paper. It is shown that in the case of an infinite chain of resonators, there are spurious solutions of the characteristic equation. These spurious solutions do not violate the correctness of direct numerical calculations, but their existence makes it difficult (or even impossible) to use approximate WKB methods for analysing chains with slowly varying parameters.

\end{abstract}

\section{Introduction}

The classical Coupled Cavity Model  (CCM) is widely used  for design of RF/microwave devices, in particular,the accelerating structures (see, for example, \cite{Bathe, Vladim, Akhiezer, Allen, Bevensee, Nishikawa1, Nishikawa2,Swain, Helm, Knapp,Swain,Yamazaki1, Gluckstern,Yamazaki2, Shintake, Gao, Sobenin, Paramonov, AyzCoupledArhive1, AyzCoupledArhive2, Kelisani} ), the travelling wave tubes (see, for example,\cite{Gilmour}), the narrow-band bandpass filters (see, for example, \cite {Hunter, Cameron, Hong}) and so on. It is also found wide application in optic and  metamaterial structures (see, for example, \cite{Nada, Opticalwaveguides}). 

These devises are based on chains of coupled resonators. Such chains belong to the class of closed structured waveguides  - waveguides that consist of similar, but not always identical, cells that couples through openings in dividing walls. Each cell couples only with two neighbouring cells. In opening structured waveguides (optic and metamaterial structures \cite{Nada, Opticalwaveguides}) each cell couples with all cells of the chain.

Despite the differences, the general approach was the same - it is necessary to find eigen modes of noncoupling cells and construct the coupling coefficients of intercoupled resonators and the external quality factors of the input and output resonators. The general coupling matrix is of importance for representing a wide range of coupled resonators  topologies. 

On the base of simple model of infinite chain of cylindrical resonators that couple  through circular openings in thing dividing walls we show that the classical CCM has problems in describing field distributions in closed structured waveguides. 

\section{Finite  Chain of Resonators. Main equations}

Consider a chain of $N_R$ cylindrical resonators with annular discs of zero thickness. The first and last resonators are connected through cylindrical openings to semi-infinite cylindrical waveguides.  We will consider only axially symmetric TM fields with  $E_z,E_r, H_\varphi$ components. Time dependence is $\exp (-i\omega {t})$. 

In each resonator we expand the electromagnetic field with the short-circuit resonant cavity modes
\begin{equation}\label{1}
{\vec E^{(k)}} = \sum\limits_q {e_q^{(k)}\,\vec E_q^{(k)}(\vec r)}  ,
\end{equation}
\begin{equation}\label{2}
{\vec H^{(k)}} = i\sum\limits_q {h_q^{(k)}\,\vec H_q^{(k)}(\vec r)} ,
 \end{equation}
 where  $q = \left\{ {0,\,m,\,n} \right\}$.
 
$\,\vec E_q^{(k)},\,\,\vec H_q^{(k)}\,$ are the solutions of homogenous Maxwell equations
\begin{equation} \label{3}
   \begin{gathered}
  rot\,\vec E_q^{(k)} = \,i\,\omega _q^{(k)}{\mu _0}\vec H_q^{(k)}\,\,, \hfill \\
  rot\,\vec H_q^{(k)} =  - i\,\omega _q^{(k)}{\varepsilon _0}\vec E_q^{(k)}\, \hfill \\ 
\end{gathered} 
\end{equation}
with boundary condition $\,{\vec E_\tau } = 0 $ on the metal surface  
\begin{equation}\label{4}
E_{m,\,n,z}^{(k)} = {J_0}\left( {\frac{{{\lambda _m}}}{{{b_k}}}r} \right)\cos \left( {\frac{\pi }{{{d_k}}}n\left( {\,z - {z_k}} \right)} \right),
  \end{equation}  
\begin{equation}\label{5}
H_{m,\,n,\,\varphi }^{(k)} =  - i\,\omega _{n,\,m}^{(k)}\frac{{{\varepsilon\varepsilon _0}\,{b_k}}}{{{\lambda _m}}}{J_1}\left( {\frac{{{\lambda _m}}}{{{b_k}}}r} \right)\cos \left( {\frac{\pi }{{{d_k}}}n\,\left( {\,z - {z_k}} \right)} \right),
\end{equation}  
\begin{equation}\label{6}
E_{m,\,n,\,r}^{(k)} = \frac{{{b_k}}}{{{\lambda _m}}}\frac{{\pi \,n}}{{{d_k}}}{J_1}\left( {\frac{{{\lambda _m}}}{{{b_k}}}r} \right)\sin \left( {\frac{\pi }{{{d_k}}}n\,\left( {\,z - {z_k}} \right)} \right),
 \end{equation} 
 \begin{equation}\label{7}
\omega _{0,\,m,\,n}^{(k)2} =\frac{c^2}{\varepsilon}  \left( {{\left( {\frac{{{\lambda _m}}}{{{b_k}}}} \right)}^2} + {{\left( {\frac{{\pi \,n}}{{{d_k}}}} \right)}^2} \right),
 \end{equation} 
 where ${J_0}({\lambda _m}) = 0$, $b_k$, $d_k$ - k-th resonator radius and length,  $a_k$ - opening radius between k-th and (k-1)-th resonators.

Amplitudes $e_q^{\left(k\right)}$ can be found if the tangential electric fields on the openings are known 
 \begin{equation}\label{8}
   \left( {\omega _q^{(k)2} - {\omega ^2}} \right)e_q^{(k)}\,\, = \frac{{i\,\omega _q^{(k)}}}{{N_q^{(k)}}}\left( {\oint\limits_{{S_k}} {[\vec E_c^{(k)}\vec H_q^{(k) * }]\,d\vec S + \oint\limits_{{S_{k + 1}}} {[\vec E_c^{(k + 1)}\vec H_q^{(k) * }]\,d\vec S} } } \right),
   \end{equation}  
\[\frac{{\omega _{0,m,n}^{(k)2}}}{{N_{n,\,m}^{(k)}}} = \frac{{2{c^2}\,\lambda _m^2}}{{{\varepsilon _0}\,\left| \varepsilon  \right|\varepsilon {\sigma _n}b_k^4\pi \,{d_k}J_1^2({\lambda _m})\,}}.\]   
In the semi-infinite waveguides the electromagnetic field can be expanded in terms of the TM eigenmodes $ \vec{\mathcal{E}}_s^{(w,p)},\,\,\vec{ \mathcal{H}}_s^{(w,p)}$ of a circular waveguide ($p=1,2$ )
\begin{equation} \label{9}
\begin{gathered}
  {{\vec H}^{(w,p)}} = \sum\limits_s {\left( {G_s^{(p)}\vec{ \mathcal{H}}_s^{(w,p)} + G_{ - s}^{(k)}\vec{ \mathcal{H}}_{ - s}^{(w,p)}} \right)}  \hfill \\
  {{\vec E}^{(w,p)}} = \sum\limits_s {\left( {G_s^{(p)}\vec{ \mathcal{E}}_s^{(w,p)} + G_s^{( - p)}\vec{ \mathcal{E}}_{ - s}^{(w,p)}} \right)}  \hfill \\ 
\end{gathered}
\end{equation}
where ${z_{w,1}} = {z_1},\,\,\,{z_{w,2}} = {z_{{N_R} + 1}}$,
\begin{equation}\label{10}
\mathcal{E}_{s,z}^{(w,p)} = {J_0}\left( {\frac{{{\lambda _s}}}{{{b_{w,p}}}}r} \right)\exp \left\{ {\gamma _s^{(w,p)}(z - {z_{w,p}})} \right\}, \hfill \\
\end{equation}
\begin{equation}\label{11}
 \mathcal{E}_{s,\,r}^{(w,p)} =  - \frac{{{b_{w,p}}}}{{{\lambda _s}}}\gamma _s^{(w,p)}{J_1}\left( {\frac{{{\lambda _s}}}{{{b_{w,p}}}}r} \right)\exp \left\{ {\gamma _s^{(w,p)}(z - {z_{w,p}})} \right\}, \hfill \\
\end{equation}
\begin{equation}\label{12}
 \mathcal{H}_{s,\varphi }^{(w,p)} =  - i\,\omega \frac{{{\varepsilon _0}{b_{w,p}}}}{{{\lambda _s}}}{J_1}\left( {\frac{{{\lambda _s}}}{{{b_{w,p}}}}r} \right)\exp \left\{ {\gamma _s^{(w,p)}(z - {z_{w,p}})} \right\}, \hfill \\ 
\end{equation}
\begin{equation}\label{13}
\gamma _s^{(w,k)2} = \frac{1}{{b_{w,k}^2}}\left( {\lambda _s^2 - \frac{{b_{w,k}^2{\omega ^2}}}{{{c^2}}}} \right).
\end{equation}

Further consideration will be based on the Moment Method. We shall use the Bessel functions as the  testing functions (${\psi _s}(x) = {J_1}({\lambda _s}x),\,\,x \in [0,1]$) and the complete set of functions that fulfil the edge condition on the diaphragm rims as the basis functions (the Meixner basis). We use such Meixner basis \cite{Eminov}
\begin{equation}\label{14}
{\varphi _s}(r) = 2\sqrt \pi  \frac{{\Gamma (s + 1)}}{{\Gamma (s - 0.5)}}\frac{1}{{\sqrt {1 - {r^2}} }}P_{2s - 1}^{ - 1}\left( {\sqrt {1 - {r^2}} } \right),
\end{equation}
where $P_n^m\left( x \right)$ are Legendre functions (or spherical functions) of the first kind,
\[{\varphi _s}\left( {\frac{r}{{{a_k}}}} \right)\xrightarrow[{r \to {a_k}}]{}\frac{{{C_s}}}{{\sqrt {1 - {{\left( {\frac{r}{{{a}}}} \right)}^2}} }}\]
 
Tangential electric fields on the openings  we expand  into such series
   \begin{equation}\label{15}
  E_r^{(k)} = \sum\limits_s {\mathcal{E}_s^{(k)}} {\varphi _s}\left( {\frac{r}{{{a}}}} \right).
   \end{equation}
 Using a procedure proposed in \cite{Amari1,Amari2,Amari3,AyzModel,AyzModelArhive}, we get such matrix equations
\begin{equation}\label{16}
   \left( {{T^{(1,1)}} - {\varepsilon ^{ - 1}}{W^{(1)}}} \right){C^{(1)}} - {T^{(1,3)}}{C^{(2)}} =  - {R^{(1,1)}}e_{010}^{(1)} + {\varepsilon ^{ - 1}}V
  \end{equation}

\begin{equation}\label{17}
     \begin{gathered}
  \left( {{T^{(k,1)}} + {T^{(k,2)}}} \right){C^{(k)}} - {T^{(k,3)}}{C^{(k + 1)}} - {T^{(k,4)}}{C^{(k - 1)}} = - {R^{(k,1)}}e_{010}^{(k)} + {R^{(k,2)}}e_{010}^{(k - 1)},\,  \hfill \\
  \,\,\,\,k = 2,3,...,{N_R} \hfill \\
   \hfill \\ 
\end{gathered}
 \end{equation} 
  
\begin{equation}\label{18}
- {T^{({N_R} + 1,4)}}{C^{({N_R})}} + \left( {{T^{({N_R} + 1,2)}} - {\varepsilon ^{ - 1}}{W^{(2)}}} \right){C^{({N_R} + 1)}} = {R^{(1{N_R} + 1,2)}}e_{010}^{({N_R}))},
\end{equation}

\begin{equation} \label{19}
  \begin{gathered}
\left( {\omega _{010}^{(k)2}- {\omega ^2}} \right)e_{010}^{(k)} = \hfill \\
   - 2\frac{{{b_k}}}{{{d_k}}}\frac{{\omega _{010}^{(k)2}}}{{\varepsilon J_1^2({\lambda _1}){\lambda _1}}}\left( { - \frac{{a_k^2}}{{b_k^2}}\sum\limits_s {C_s^{(k)}{j_{2s - 1}}\left( {\frac{{{\lambda _1}{a_k}}}{{{b_k}}}} \right)}  + \frac{{a_{k + 1}^2}}{{b_k^2}}\sum\limits_s {C_s^{(k + 1)}{j_{2s - 1}}\left( {\frac{{{\lambda _1}{a_{k + 1}}}}{{{b_k}}}} \right)} } \right), \hfill \\
  k = 1,2,...,{N_R}. \hfill \\ 
\end{gathered} 
  \end{equation} 
Matrices  $T, \,W$ and vector $R$ are defined in Appendix 1. 
We  introduce the fundamental solution of system \eqref{16} - \eqref{18} as the solution of such difference matrix  equations
\begin{equation}\label{20}
  \begin{gathered}
  \left( {{T^{(1,1)}} - {\varepsilon ^{ - 1}}{W^{(1)}}} \right){Y^{(1,{k_0})}} - {T^{(1,3)}}{Y^{(2,{k_0})}} =  - {R^{(1,1)}}{\delta _{1,{k_0}}}, \hfill \\
  \left( {{T^{(k,1)}} + {T^{(k,2)}}} \right){Y^{(k,{k_0})}} - {T^{(k,3)}}{Y^{(k + 1,{k_0})}} - {T^{(k,4)}}{Y^{(k - 1,{k_0})}} =\hfill \\  - {R^{(k,1)}}{\delta _{k,{k_0}}} + {R^{(k,2)}}{\delta _{k - 1,{k_0}}}, 
  k = 2,3,...,{N_R}, \hfill \\
   - {T^{({N_R} + 1,4)}}{Y^{({N_R},{k_0})}} + \left( {{T^{({N_R} + 1,2)}} - {\varepsilon ^{ - 1}}{W^{(2)}}} \right){Y^{({N_R} + 1,{k_0})}} = {R^{({N_R} + 1,2)}}{\delta _{{N_R},{k_0}}}, \hfill \\ 
\end{gathered} 
\end{equation}

\begin{equation}\label{21}
\begin{gathered}
  \left( {{T^{(1,1)}} - {\varepsilon ^{ - 1}}{W^{(1)}}} \right){Y^{(1)}} - {T^{(1,3)}}{Y^{(2)}} =  = {\varepsilon ^{ - 1}}V, \hfill \\
  \left( {{T^{(k,1)}} + {T^{(k,2)}}} \right){Y^{(k)}} - {T^{(k,3)}}{Y^{(k + 1)}} - {T^{(k,4)}}{Y^{(k - 1)}} = 0, \hfill \\
  k = 2,3,...,{N_R}, \hfill \\
   - {T^{({N_R} + 1,4)}}{Y^{({N_R})}} + \left( {{T^{({N_R} + 1,2)}} - {\varepsilon ^{ - 1}}{W^{(2)}}} \right){Y^{({N_R} + 1)}} = 0 \hfill \\ 
\end{gathered} 
\end{equation}
Then the solution of \eqref{16} - \eqref{18} is written as 
\begin{equation}\label{22}
C_s^{(k)} = \sum\limits_{{k_0} = 1}^{{N_R}} {Y_s^{(k,{k_0})}e_{010}^{({k_0})}}  + Y_s^{(k)}
\end{equation}
Its substitution in \eqref{19} gives the basic CCM equation
\begin{equation}\label{23}
\left( {\omega _{010}^{(k)2} - {\omega ^2}} \right)e_{010}^{(k)}\,\, = \omega _{010}^{(k)2}\sum\limits_{{k_0} = 1}^{{N_R}} {{\alpha ^{(k,{k_0})}}e_{010}^{({k_0})}}  + \omega _{010}^{(k)2}{F^{(k)}},
\end{equation}
where
\begin{equation}\label{24}
{\alpha ^{(k,{k_0})}} =  - 2\frac{{{b_k}}}{{{d_k}J_1^2({\lambda _1}){\lambda _1}}}\left( { - \frac{{a_k^2}}{{b_k^2}}\sum\limits_s {Y_s^{(k,{k_0})}{j_{2s - 1}}\left( {\frac{{{\lambda _1}{a_k}}}{{{b_k}}}} \right)}  + \frac{{a_{k + 1}^2}}{{b_k^2}}\sum\limits_s {Y_s^{(k + 1,{k_0})}{j_{2s - 1}}\left( {\frac{{{\lambda _1}{a_{k + 1}}}}{{{b_k}}}} \right)} } \right)
\end{equation}
\begin{equation}\label{25}
{F^{(k)}} =  - 2\frac{{{b_k}}}{{{d_k}J_1^2({\lambda _1}){\lambda _1}}}\left( { - \frac{{a_k^2}}{{b_k^2}}\sum\limits_s {Y_s^{(k)}{j_{2s - 1}}\left( {\frac{{{\lambda _1}{a_k}}}{{{b_k}}}} \right)}  + \frac{{a_{k + 1}^2}}{{b_k^2}}\sum\limits_s {Y_s^{(k + 1)}{j_{2s - 1}}\left( {\frac{{{\lambda _1}{a_{k + 1}}}}{{{b_k}}}} \right)} } \right)
\end{equation}
Mode amplitudes in cylindrical waveguides can be found using the following formulas
\begin{equation}\label{26}
 \begin{gathered}
  G_{ - 1}^{(1)} = 1 + 2\frac{{a_1^2{\lambda _1}}}{{J_1^2\left( {{\lambda _s}} \right)b_{w,1}^2\gamma _1^{(1)}{b_{w,1}}}}\sum\limits_{s'} {C_{s'}^{(1)}{j_{2s' - 1}}\left( {\frac{{{\lambda _1}{a_1}}}{{{b_{w,1}}}}} \right)}  \hfill \\
  G_{ - s}^{(1)} = 2\frac{{a_1^2{\lambda _s}}}{{J_1^2\left( {{\lambda _s}} \right)b_{w,1}^2\gamma _s^{(1)}{b_{w,1}}}}\sum\limits_{s'} {C_{s'}^{(1)}{j_{2s' - 1}}\left( {\frac{{{\lambda _s}{a_1}}}{{{b_{w,1}}}}} \right)} ,\,\,\,\,s = 2,3,... \hfill \\ 
\end{gathered} 
\end{equation}

\begin{equation}\label{27}
G_s^{(2)} =  - 2\frac{{a_{{N_R} + 1}^2{\lambda _s}}}{{J_1^2\left( {{\lambda _s}} \right)b_{w,2}^2\gamma _s^{(2)}{b_{w,2}}}}\sum\limits_{s'} {C_{s'}^{({N_R} + 1)}{j_{2s' - 1}}\left( {\frac{{{\lambda _s}{a_{{N_R} + 1}}}}{{{b_{w,2}}}}} \right)} ,\,\,s = 1,2,...
\end{equation}
 
 The model described above has no limitations. We get the exact coupling matrix \eqref{24} and the vector of distributed external sources \eqref{25}.
\section{Infinitive  Chain of Resonators}

Consider an infinitive (${N_R}\rightarrow\infty$) homogeneous ($a_k=a,\,b_k=b,\,d_k=d$) chain. Equations \eqref{17},\eqref{19} we write as
 \begin{equation}\label{28}
 e_{010}^{(k)}\,\, =  - q\,\sum\limits_{s = 1}^\infty  {{R_s}\left( {\mathcal{E}_s^{(k)} - \mathcal{E}_s^{(k + 1)}} \right)}, 
 \end{equation}
\begin{equation}\label{29}
 2{T^{(1)}}{\mathcal{E}^{(k)}} - {T^{(2)}}\,{\mathcal{E}^{(k + 1)}} - {T^{(2)}}{\mathcal{E}^{(k - 1)}} = R\left( {e_{010}^{(k)} - e_{010}^{(k - 1)}} \right),
 \end{equation} 
where  
\begin{equation}\label{30}
q = \frac{{\omega _{010}^2}}{{\left( {\omega _{010}^2 - {\omega ^2}} \right)}}\frac{{\,{a^3}}}{{{b^2}d\,\,J_1^2({\lambda _1})}},
\end{equation} 
\begin{equation}\label{31}
{R_s} =\frac{1}{2} \frac{{{J_0}\left( {\frac{{a{\lambda _1}}}{b}} \right)}}{{{{\left( {\frac{{a{\lambda _1}}}{b}} \right)}^2} - {{\left( {{\lambda _s}} \right)}^2}}}
 \end{equation} 

\subsection{ Systems of matrix equations }
 
Substitution \eqref{28}  into \eqref{29} gives 
\begin{equation}\label{32}
2{\bar T^{(1)}}{\mathcal{E}^{(k)}} - {\bar T^{(3)}}\,{\mathcal{E}^{(k + 1)}} - {\bar T^{(3)}}{\mathcal{E}^{(k - 1)}} = 0,
\end{equation} 
where
\begin{equation}\label{33}
\bar T_{s,s'}^{(1,2)} = \frac{{a^2}}{{b^2}}\sum\limits_m {\bar\Lambda _m^{(1,2)}} \frac{{{\lambda _m}}}{{J_1^2({\lambda _m})}}\frac{{{J_0}\left( {\frac{{{a}{\lambda _m}}}{{{b}}}} \right){j_{2s - 1}}\left( {\frac{{{\lambda _m}{a}}}{{{b}}}} \right)}}{{{{\left( {\frac{{{a}{\lambda _m}}}{{{b}}}} \right)}^2} - \lambda _{s'}^2}},
\end{equation}
\begin{equation}\label{34}
\begin{gathered}
  \bar \Lambda _m^{(1)} = \frac{{d\,\cosh(d{h_m})}}{{bd{h_m}\,\sinh(d{h_m})}}, \hfill \\
  \bar \Lambda _m^{(2)} = \frac{d}{{bd{h_m}\,\sinh(d{h_m})}}. \hfill \\ 
\end{gathered} 
\end{equation}
It can be shown that the matrix $T^{(2)}$ is invertible and the equation \eqref{32} can be rewritten  as 
\begin{equation}\label{35}
\bar T{\mathcal{E}^{(k)}} = \,{\mathcal{E}^{(k + 1)}} + {\mathcal{E}^{(k - 1)}},
\end{equation}
where
\begin{equation}\label{36}
\bar T = 2{\bar T^{(2) - 1}}{\bar T^{(1)}}.
\end{equation} 
The general solution of the  difference equation \eqref{32} is \cite{AyzModel,AyzModelArhive}
\begin{equation}\label{37}
{\mathcal{E}^{(k)}} = \sum\limits_{s = 1}^\infty  {\bar C_s^{(1)}\bar \lambda _s^{(1)k}{{\bar U}_s}}  + \sum\limits_{s = 1}^\infty  {\bar C_s^{(2)}\bar \lambda _s^{(2)k}{{\bar U}_s}} 
\end{equation}
where
\begin{equation}\label{38}
\bar T{U_s} = {\bar \mu _s}{\bar U_s},
\end{equation}
\begin{equation}\label{39}
\begin{gathered}
  \bar \lambda _s^2 - {{\bar \mu }_s}{{\bar \lambda }_s} + 1 = 0, \hfill \\
  \bar \lambda _s^{(1,2)} = \frac{{{{\bar \mu }_s}}}{2} \pm \frac{1}{2}\sqrt {\bar \mu _s^2 - 4} , \hfill \\ 
\end{gathered}
\end{equation}
\begin{equation}\label{40}
\begin{gathered}
  {{\bar \mu }_s} = \bar \lambda _s^{(1)} + \bar \lambda _s^{(2)}, \hfill \\
  \bar \lambda _s^{(1)}\bar \lambda _s^{(2)} = 1. \hfill \\ 
\end{gathered}
\end{equation}

We used EVCRG programm from IMSL Fortran Numerical Library to calculate $\bar \lambda _s$. Results are presented in Tab.1 ( first and second columns, angles are given in degrees).
It can be seen that chosen frequency f=2.856 GHz lay in the first passband ($E_{01}$) and others modes ($E_{02}, E_{03}...$) are evanescent with zero angles.  These results are fitted well the existing theories of periodic waveguides. For the frequency that lay in the first propagation zone  we have two propagating waves and the infinitive number of evanescent waves with characteristic multipliers with zero phases. Between the first and the second propagation zones there are only evanescent oscillations with the phase shift per cell equals $\pi$. The other  evanescent oscillations have zero phase shift. At the end of the second stop band  evanescent $E_{01}$  wave with $\pi$-shift transforms into propagating   $E_{02}$ wave with $\pi$-shift too \cite{AyzEvanescent1}. Therefore, in the second passband $E_{02}$ wave has negative dispersion. 
\begin{quote}
\centering
\begin{tabular}{|c|c|c|c|}  \hline
\multicolumn {4}{|c|} {Table 1 Calculated values of $ \bar \lambda _s$ } \\   \hline
\multicolumn {4}{|c|}{f=2.856 GHz, a=0.99 cm, b=4.08896 cm, d=3.4989 cm } \\   \hline
{Matrix  $4\times4$}&{Matrix  $3\times3$}& {CCM (9)} & {CCM(7) } \\  \hline
3.85E+08$\angle0^{\circ}$&	\,  &
194.06$\angle0^{\circ}$  & \, \\ \hline
0.00E+00$\angle0^{\circ}$&	\,&\
5.15E-03$\angle0^{\circ}$\, \\ \hline

1.62E+06$\angle0^{\circ}$&1.64E+06$\angle0^{\circ}$&	
193.67$\angle-89.96^{\circ}$&194.06$\angle-59.96^{\circ}$ \\ \hline
6.18E-07$\angle0^{\circ}$&	6.18E-07$\angle0^{\circ}$&
5.16E-03$\angle89.96^{\circ}$&	5.15E-03$\angle59.96^{\circ}$ \\ \hline

6.09E+03 $\angle0^{\circ}$&6.09E+03 $\angle0^{\circ}$&
193.67$\angle-89.96^{\circ}$&194.06$\angle-59.96^{\circ}$ \\ \hline
1.64E-04$\angle0^{\circ}$&1.64E-04$\angle0^{\circ}$&	
5.16E-03$\angle89.96^{\circ}$&	5.16E-03$\angle59.96^{\circ}$\\ \hline

1.0$\angle-119.994^{\circ}$&1.0$\angle-119.994^{\circ}$&1.0$\angle-119.997^{\circ}$	&	1.0$\angle-119.997^{\circ}$\\ \hline
1.0$\angle119.994^{\circ}$&1.0$\angle119.994^{\circ}$&1.0$\angle199.997^{\circ}$&	1.0$\angle119.997^{\circ}$\\ \hline
\end{tabular} 
\end{quote}

\subsection{ Coupled Cavity Model }

The equation \eqref{29} we rewrite as
\begin{equation}\label{41}
2{T^{(1)}}{\mathcal{E}^{(k)}} - {T^{(3)}}\,{\mathcal{E}^{(k + 1)}} - {T^{(3)}}{\mathcal{E}^{(k - 1)}} = R\sum\limits_{j =  - \infty }^\infty  {\left( {e_{010}^{(j)} - e_{010}^{(j - 1)}} \right){\delta _{k,j}}}  = R\sum\limits_{j =  - \infty }^\infty  {e_{010}^{(j)}\left( {{\delta _{k,j}} - {\delta _{k,j + 1}}} \right)}.
\end{equation}

Let's introduce the fundamental solution  as the solution of such difference equation ( $i$ - the number of diaphragm, $j$  - the number of cell with a source)
\begin{equation}\label{42}
2{T^{(1)}}{Y^{(i,j)}} - {T^{(3)}}{Y^{(i + 1,j)}} - {T^{(3)}}{Y^{(i - 1,j)}} = R{\delta _{i,j}} 
\end{equation}

We will suppose that eigenvalues ${\mu _s}$ of the matrix $ T = 2{T^{(2) - 1}}{T^{(1)}}$  ($T{U_l} = {\mu _l}{U_l}$) take such values that the quantities 
\begin{equation}\label{43}
\rho _s^{(1,2)} = \frac{{{\mu _s}}}{2} \pm \frac{1}{2}\sqrt {\mu _s^2 - 4}
\end{equation}
are the real numbers. In this case, we can impose the condition at infinity
\begin{equation}\label{44}
\left| {{Y^{(i,j)}}} \right|\xrightarrow[{i \to  \pm \infty }]{}0
\end{equation}
and the solution of \eqref{26} we can write as
\begin{equation}\label{45}
{Y^{(g,0)}} = \sum\limits_{l = 1}^\infty {A_l}{U_l}\left\{ \begin{gathered}
  \lambda _l^{(1)g},\,\,\,g < 0, \hfill \\
  \lambda _l^{(2)g},\,\,\,g \geqslant 0, \hfill \\ 
\end{gathered}  \right. 
\end{equation}
where
\begin{equation}\label{46}
\begin{gathered}
  {A_l} = \frac{{{r_l}}}{{\lambda _l^{(2)} - \lambda _l^{(1)}}}, \hfill \\
  R = \sum\limits_{l = 1}^\infty  {{r_l}{U_l}} . \hfill \\ 
\end{gathered} 
\end{equation}
The solution of equation \eqref{41} can be written as
\begin{equation}\label{47}
\mathcal{E}_s^{(k)} = \sum\limits_{j =  - \infty }^{j =  - \infty } {\left( {Y_s^{(k,j)} - Y_s^{(k - 1,j)}} \right)e_{010}^{(j)}} 
\end{equation}
and the equation \eqref{28} takes the form
\begin{equation}\label{48}
\begin{gathered}
  e_{010}^{(k)}\,\, =  - q\,\sum\limits_{j =  - \infty }^\infty  {e_{010}^{(j)}\sum\limits_{s = 1}^\infty  {{R_s}\left( {2Y_s^{(k,j)} - Y_s^{(k - 1,j)} - Y_s^{(k + 1,j)}} \right)} } \; =  \hfill \\
   =  - q\,\sum\limits_{j =  - \infty }^\infty  {e_{010}^{(k + j)}\sum\limits_{s = 1}^\infty  {{R_s}\left( {2Y_s^{(k,j + k)} - Y_s^{(k - 1,j + k)} - Y_s^{(k + 1,j + k)}} \right)} }.  \hfill \\ 
\end{gathered}
\end{equation}
From \eqref{45} it follows that ${Y^{(i,j + g)}} = {Y^{(i - g,j)}}$, $
  {Y^{(i,0)}} = {Y^{( - i,0)}}$ and we get the final equation of CCM
\begin{equation}\label{49}
\left( {\omega _{010}^2 - {\omega ^2}} \right)e_{010}^{(k)} =  - \omega _{010}^2\sum\limits_{k_0 =  - \infty }^\infty  {e_{010}^{(k + k_0)}{\alpha ^{\left(k_0 \right)}}} 
\end{equation}
where
\begin{equation}\label{50}
\alpha^{\left(k_0 \right)}\left( \omega \right) = \frac{{{a^3}}}{{{b^2}d\,J_1^2({\lambda _1})}}\sum\limits_{s = 1}^\infty  {{R_s}\left( {2Y_s^{(k_0,0)} - Y_s^{(k_0 - 1,0)} - Y_s^{(k_0 + 1,0)}}. \right)} 
\end{equation}

It is often assumed that we can neglect "long couplings"
\begin{equation}\label{51}
\left( {\omega _{010}^2 - {\omega ^2}} \right)e_{010}^{(k)} =  - \omega _{010}^2\sum\limits_{k_0 =  - N}^N {e_{010}^{(k + k_0)}{\alpha^{\left(k_0 \right)}}} ,
\end{equation}
It was also  assumed  \cite {AyzCoupledArhive1,AyzCoupledArhive2}
that the sum in \eqref{50} can be also truncated 
\begin{equation}\label{52}
{\alpha^{\left(k_0 \right)}} = \frac{{{a^3}}}{{{b^2}d\,J_1^2({\lambda _1})}}\sum\limits_{s = 1}^S {{R_s}\left( {2Y_s^{k_0j,0)} - Y_s^{(k_0 - 1,0)} - Y_s^{(k_0 + 1,0)}} \right)},
\end{equation}
\begin{equation}\label{53}
\begin{gathered}
  2T_{s',s}^{(1)}Y_s^{(i,0)} - T_{s',s}^{(2)}Y_s^{(i + 1,0)} - T_{s',s}^{(2)}Y_s^{(i - 1,0)} = {R_{s'}}{\delta _{i,0}},\,\, \hfill \\
  i =  - I + 1,...,0,...I - 1;\,\,\,I > N,\,\,\,\,s,s' = 1,...,S, \hfill \\
  2T_{s',s}^{(1)}Y_s^{( - I,0)} - T_{s',s}^{(2)}Y_s^{( - I + 1,0)} = 0, \hfill \\
  2T_{s',s}^{(1)}Y_s^{(I,0)} - T_{s',s}^{(2)}Y_s^{(I - 1,0)} = 0. \hfill \\ 
\end{gathered} 
\end{equation}

Equation \eqref{51} is widely used for description of different objects. 
We made assumption that $\rho_s^{(1,2)}$ are the real numbers. Tab.2 shows that it is true.
\begin{quote}
\centering
\begin{tabular}{|c|c|c|}  \hline
\multicolumn {3}{|c|} {Table 2 Calculated values of $\mu_s$ and $  \rho_s$ } \\   \hline
\multicolumn {3}{|c|}{{f=2.856 GHz, a=0.99 cm, b=4.08896 cm, d=3.4989 cm }}\\   \hline
\multicolumn {3}{|c|}{ I=6, N=5, S=5 }\\   \hline
{$\mu_s$}&{ $\rho_s^{(2)}$} & {$\rho_s^{(1)}$ }  \\  \hline
(-1.93E+02,0.0E+0)&	5.17E-03& 1.93E+02 \\ \hline
(2.16E+04,0.0E+0)&4.61E-05	&2.16E+04\\ \hline
(2.02E+07,0.0E+0)&4.93E-08	&2.02E+07 \\ \hline
(1.26E+10,0.0E+0)&7.91E-11&1.26E+10 \\ \hline
(5.32E+12,0.0E+0)&	1.88E-13& 5.32E+12\\ \hline
\end{tabular} 
\end{quote}

If we seek the solution of \eqref{51} as $e_{010}^{(k)}\sim\lambda^k$, we get the characteristic equation
\begin{equation}\label{54}
\sum\limits_{{k_0} =  - N}^N {\lambda ^{k_0}{\alpha ^{\left(k_0 \right)}\left( \omega \right)}}  + \frac{{\left( {\omega _{010}^2 - {\omega ^2}} \right)}}{{\omega _{010}^2}} = 0.
\end{equation}

Solutions of this equation are given in Table 1 (third and fourth columns) for different N. We see that propagating $E_{01}$ wave has the same phase shift for different approaches, but other solutions are very different. We obtain solutions that represent "non-physical" evanescent waves with phase shifts that differ from 0 and $\pi$.  Amplitudes of these solutions are nearly the same, but phases strongly depend on the number of interacting oscillators. 

To understand the reasons for such solutions, we consider the simplest case when the tangential electric field is described by one coefficient  
 \begin{equation}\label{55}
   E_r^{(k)} = { \mathcal{E}_1^{(k)}}\varphi_1\left(r/a\right).
 \end{equation}
Then \eqref{28},\eqref{29} transform into
\begin{equation}\label{56}
e_{010}^{(k)}\,\, =  - q{R_1}\left( {\mathcal{E}_1^{(k)} - \mathcal{E}_1^{(k + 1)}}, \right)\
\end{equation}
\begin{equation}\label{57}
{T^{(E)}}\mathcal{E}_1^{(k)} = {R_1}\left( {e_{010}^{(k)} - e_{010}^{(k - 1)}} \right),
\end{equation}
where operator ${T^{(E)}} = 2T_{11}^{(1)} - T_{11}^{(3)}{\sigma ^{( + )}}\, - T_{11}^{(3)}{\sigma ^{( - )}}$, ${\sigma ^{( + )}},\,\,{\sigma ^{( - )}}$ - shift operators (${\sigma ^{( + )}}{x^{(k)}} = {x^{(k + 1)}}$, ${\sigma ^{( - )}}{x^{(k)}} = {x^{(k - 1)}}$).

Substitution \eqref{56} into \eqref{57} gives following equation
\begin{equation}\label{58}
2\left( {T_{11}^{(1)} + qR_1^2} \right)\mathcal{E}_1^{(k)} - \left( {T_{11}^{(3)} + qR_1^2} \right)\mathcal{E}_1^{(k + 1)}\, - \left( {T_{11}^{(3)} + qR_1^2} \right)\mathcal{E}_1^{(k - 1)} = 0.
\end{equation}
The solution of this equation is
\begin{equation}\label{59}
\mathcal{E}_1^{(k)} = {C_1}\lambda _1^k + {C_2}\lambda _2^k,
\end{equation}
where $\lambda_1$ and $\lambda_2$ are the solutions of the characteristic equation
\begin{equation}\label{60}
{\lambda ^2} - 2\frac{{\left( {T_{11}^{(1)} + qR_1^2} \right)}}{{\left( {T_{11}^{(3)} + qR_1^2} \right)}}\lambda  + 1 = 0.
\end{equation}
Solutions to this equation are (f=2.856 GHz, a=0.99 cm, b=4.08896 m, d=3.4989 cm):
 $ \lambda_1=1{\angle}119.41^{\circ}$, $ \lambda_2=1{\angle}-119.41^{\circ}$ ( compare with results in Table 1). 

We can also apply the operator $T^{(E)}$ to the right and left sides of equation \eqref{56} and get the equation like \eqref{58}
\begin{equation} \label{61}
2\left( {T_{11}^{(1)} + qR_1^2} \right)e_{010}^{(k)} - \left( {T_{11}^{(3)} + qR_1^2} \right)e_{010}^{(k + 1)}\, - \left( {T_{11}^{(3)} + qR_1^2} \right)e_{010}^{(k - 1)} = 0.
\end{equation}

The solution of equation \eqref{57} can be also written  using the solution $Y_1^{(k,j)}$ of  the fundamental equation 
\begin{equation}\label{62}
2T_{11}^{(1)}Y_1^{(k,j)} - T_{11}^{(3)}Y_1^{(k + 1,j)}\, - T_{11}^{(3)}Y_1^{(k - 1,j)} = {R_1}{\delta _{k,j}}.
\end{equation}
If $\left| {{\chi _2}} \right| < 1$, where
\begin{equation}\label{63}
{\chi _{1,2}} = \frac{{T_{11}^{(1)}}}{{T_{11}^{(3)}}} \pm \sqrt {{{\left( {\frac{{T_{11}^{(1)}}}{{T_{11}^{(3)}}}} \right)}^2} - 1},
\end{equation}
then the solution of \eqref{62} is
\begin{equation}\label{64}
Y_1^{(k,j)} = \left\{ \begin{gathered}
  A\chi _1^{k - j},\,k - j < 0\,\,, \hfill \\
  A\chi _2^{k - j},\,k - j \geqslant 0, \hfill \\ 
\end{gathered}  \right.
\end{equation}
\begin{equation}\label{65}
A = \frac{{{R_1}}}{{2\left( {T_{11}^{(1)} - {\chi _2}T_{11}^{(3)}} \right)}}.
\end{equation}
     
The solution of \eqref{57} is
\begin{equation}\label{66}
\mathcal{E}_1^{(k)} = \sum\limits_j {Y_1^{(k,j)}\left( {e_{010}^{(j)} - e_{010}^{(j - 1)}} \right)}.
\end{equation}
Substitution \eqref{66} into \eqref{56} gives the final equation
\begin{equation}\label{67}
\begin{gathered}
  e_{010}^{(k)}\,\, =  - q{R_1}\sum\limits_{j =  - \infty }^\infty  {e_{010}^{(k + j)}\left( {2Y_1^{(k - j,k)} - Y_1^{(k - j - 1,k)} - Y_1^{(k - j + 1,k)}} \right)}  =  \hfill \\
   =  - q{R_1}Ae_{010}^{(k)}2\left( {1 - {\chi _2}} \right) - q{R_1}A\sum\limits_{j = 1}^\infty  {\left( {e_{010}^{(k + j)} + e_{010}^{(k - j)}} \right)\left( {2\chi _2^j - \chi _2^{j + 1} - \chi _2^{j - 1}} \right)}  =  \hfill \\
   =  - q{R_1}2Ae_{010}^{(k)}\left( {1 - {\chi _2}} \right) - q{R_1}A\left( {2 - {\chi _2} - {\chi _1}} \right)\sum\limits_{j = 1}^\infty  {\chi _2^j\left( {e_{010}^{(k + j)} + e_{010}^{(k - j)}} \right)}.  \hfill \\ 
\end{gathered}
\end{equation}
The characteristic equation of the difference equation \eqref{67} coincide 
with the equation 
\begin{equation}\label{68}
1 + q{R_1}2A\left( {1 - {\chi _2}} \right) - q{R_1}A\frac{{{{\left( {1 - {\chi _2}} \right)}^2}}}{{{\chi _2}}}\sum\limits_{j = 1}^N {\left( {{\lambda ^j} + {\lambda ^{ - j}}} \right)\chi _2^j}  = 0,
\end{equation}
when $N\rightarrow\infty$.

The equation \eqref{68} we can rewrite as
\begin{equation}\label{69}
  \begin{gathered}
  {\lambda ^2} - 2\frac{{\left[ {T_{11}^{(1)} + qR_1^2} \right]}}{{\left( {T_{11}^{(3)} + qR_1^2} \right)}}\lambda  + 1 =  - qR_1^2\frac{{\left( {1 - {\chi _2}} \right)}}{{\left( {1 + {\chi _2}} \right)}} \times  \hfill \\
  \frac{{\left[ {\left( {{\chi _2} - \lambda } \right)\lambda {{\left( {\lambda {\chi _2}} \right)}^N} + \left( {\lambda {\chi _2} - 1} \right){{\left( {{\chi _2}/\lambda } \right)}^N}} \right]}}{{\left( {T_{11}^{(3)} + qR_1^2} \right)}}. \hfill \\ 
\end{gathered} 
\end{equation}

If we suppose that 
\begin{equation}\label{70}
\left| {\lambda {\chi _2}} \right| < 1,\,\,\left| {{\chi _2}/\lambda } \right| < 1
\end{equation}
and the number N tends to infinity, the right hand side of the equation \eqref{69} tends to zero and this 2(N+1) order characteristic  equation transforms into the two order equation that coincide with the equation \eqref{60}.  

The equation \eqref{69} under the conditions \eqref{70} has two solutions very close to the solutions of the equation \eqref{60} , and 2N additional solutions.

For considered above parameter values f=2.856 GHz, a=0.99 cm, b=4.08896 cm, d=3.4989 cm  ${\chi _2}$ = 5.03E-003. Therefore, conditions \eqref{70} are fulfilled in passband ($\left| \lambda\right|=1)$ and even in some part of stopband. 

A distinctive feature of the difference equation  \eqref{67} is that  despite the presence of an infinite sum (infinite order), its characteristic equation is of the second order (at least in some frequency domain).

The above consideration clarifies the reason for the appearance of the spurious solutions in the main equation of the Coupled Cavity Model \eqref{51} - this is a consequence of the truncation of the infinite sum.

It is well known that without the evanescent eigenfields  we cannot get the correct characteristics of inhomogeneities in waveguides. 
Since we cannot operate numerically with infinite sums, using this approach for study non-homogeneous infinite chains may become problematic because the CCM does not correctly describe the evanescent eigenfields. 

To clarify the correctness of using CCM to study infinite inhomogeneous resonator chains it is necessary to compare the values of the coupling coefficients calculated on the basis of the approximate equations \eqref{52}-\eqref{53} with the  values obtained using the exact equations \eqref{16} - \eqref{24}.  
The number of resonators in the finite chain must be sufficient to exclude the influence of boundaries on the coupling coefficients of cells in the middle of the chain.
We have made a series of calculations for different inhomogeneities inside the finite chain and the same non-uniform insertions in the infinitive chain.  Their results show that there is such number $N$ in \eqref{53} that the coupling coefficients calculated on the basis of approaches \eqref{52} and \eqref{24} are almost the same. 

Therefore, the use of the CCM for the numerical study of the characteristics of infinitive resonator chains gives correct results.
The model of an infinite chain of resonators plays an important role in calculating the electrodynamic characteristics of inhomogeneities in structured waveguides. Indeed, based on the CCM of an infinitive chain, one can study the processes of wave propagation in inhomogeneous resonator chains. This can be done if we assume that there are homogeneous fragments before and after inhomogeneities. In this case, in homogeneous fragments at a sufficient distance from the interfaces (when all evanescent waves are damped), one can look for amplitudes in the form
\begin{equation} \label{71}
e_{010}^{(k)}\, = \left\{ {\begin{array}{*{20}{c}}
  {\exp \left\{ {i{\varphi _1}\left( {k - {k_1}} \right)} \right\} + R\exp \left\{ { - i{\varphi _1}\left( {k - {k_1}} \right)} \right\},\,\,k < {k_1},} \\ 
  {T\exp \left\{ {i{\varphi _2}\left( {k - {k_2}} \right)} \right\}\,\,\,\,\,\,\,\,\,\,\,\,\,\,\,\,\,\,\,\,\,\,\,\,\,\,\,\,\,\,\,\,\,\,\,\,\,\,\,\,\,\,\,\,\,\,\,\,\,k > {k_2},} 
\end{array}} \right.
\end{equation}
where $R$ and $T$ are the reflection and transition coefficients. This approach avoids the use of an ideal load or tuned couplers. This is especially useful when we are studying frequency dependencies.

But existence of spurious solutions make it difficult (or even impossible) to use the approximate methods for analysing the chains with slow varying parameters (WKB approach). As shown by \cite{AyzTrans}, in order to obtain the WKB equations from the original difference equation, it is necessary to know the local characteristic multipliers.  It is unknown how the spurious solutions will influence on WKB equations, since their number can be great. But from the physical point of view, including the spurious solutions in consideration will be wrong.

It is necessary to use other approaches with correct eigenmode basis for analysing the chains with slow varying parameters \cite{AyzModelArhive,AyzInhomogeneousArhive,AyzModif}.
   
\section{Conclusions}
Summarizing the above, we can draw the following conclusions:

1. We can use the CCM to numerical study the characteristics of infinitive inhomogeneous resonator chains.

2. We cannot use the CCM to obtain   approximate equations (the WKB approach) for the analysis of chains with slowly varying parameters. 

\section{Appendix 1}
   
 \[ T_{s,s'}^{(k,1)} = \frac{{a_k^2}}{{b_k^2\,}}\sum\limits_m {\Lambda_m^{(1,k)}} \frac{{{\lambda _m}}}{{J_1^2({\lambda _m})}}\,\frac{{{j_{2s' - 1}}\left( {\frac{{{\lambda _m}{a_k}}}{{{b_k}}}} \right){J_0}\left( {\frac{{{a_k}{\lambda _m}}}{{{b_k}}}} \right)}}{{{{\left( {\frac{{{a_k}{\lambda _m}}}{{{b_k}}}} \right)}^2} - {{\left( {{\lambda _s}} \right)}^2}}}\]

\[T_{s,s'}^{(k,2)} = \frac{{a_k^2}}{{b_{k - 1}^2}}\sum\limits_m 
{\Lambda _m^{(1,k - 1)}} \frac{{{\lambda _m}}}{{J_1^2({\lambda _m})}}\,\,\frac{{{j_{2s' - 1}}\left( {\frac{{{\lambda _m}{a_k}}}{{{b_{k - 1}}}}} \right){J_0}\left( {\frac{{{a_k}{\lambda _m}}}{{{b_{k - 1}}}}} \right)}}{{{{\left( {\frac{{{a_k}{\lambda _m}}}{{{b_{k - 1}}}}} \right)}^2} - {{\left( {{\lambda _s}} \right)}^2}}}\]

\[T_{s,s'}^{(k,3)} = \frac{{a_{k + 1}^2}}{{b_k^2}}\sum\limits_m {\Lambda _m^{(2,k)}} \frac{{{\lambda _m}}}{{J_1^2({\lambda _m})}}\,\,\frac{{{j_{2s' - 1}}\left( {\frac{{{\lambda _m}{a_{k + 1}}}}{{{b_k}}}} \right){J_0}\left( {\frac{{{a_k}{\lambda _m}}}{{{b_k}}}} \right)}}{{{{\left( {\frac{{{a_k}{\lambda _m}}}{{{b_k}}}} \right)}^2} - {{\left( {{\lambda _s}} \right)}^2}}}\]

\[T_{s,s'}^{(k,4)} = \frac{{a_{k - 1}^2}}{{b_{k - 1}^2}}\sum\limits_m {\Lambda _m^{(2,k - 1)}} \frac{{{\lambda _m}}}{{J_1^2({\lambda _m})}}\,\frac{{{j_{2s' - 1}}\left( {\frac{{{\lambda _m}{a_{k - 1}}}}{{{b_{k - 1}}}}} \right){J_0}\left( {\frac{{{a_k}{\lambda _m}}}{{{b_{k - 1}}}}} \right)}}{{{{\left( {\frac{{{a_k}{\lambda _m}}}{{{b_{k - 1}}}}} \right)}^2} - {{\left( {{\lambda _s}} \right)}^2}}}\]

 \[\Lambda _m^{(1,k)} = \left\{ {\begin{array}{*{20}{c}}
  {\frac{{{d_k}\cosh({d_k}h_m^{(k)})}}{{{b_k}{d_k}h_m^{(k)}\sinh({d_k}h_m^{(k)})}} - \frac{{{d_k}}}{{{b_k}}}\frac{1}{{\left( {{\pi ^2} + d_k^2h_{\,m}^{(k)2}} \right)}},\,\,\,m = 1,} \\ 
  {\frac{{{d_k}\cosh({d_k}h_m^{(k)})}}{{{b_k}{d_k}h_m^{(k)}\sinh({d_k}h_m^{(k)})}},\,\,\,\,\,\,\,\,\,\,\,\,\,\,\,\,\,\,\,\,\,\,\,\,\,\,\,\,\,\,\,\,\,\,\,\,\,\,\,\,\,\,\,\,\,\,m \ne 1,} 
\end{array}} \right.\] 

\[\Lambda _m^{(2,k)} = \left\{ {\begin{array}{*{20}{c}}
  {\frac{{{d_k}}}{{{b_k}{d_k}h_m^{(k)}\sinh({d_k}h_m^{(k)})}} - \frac{{{d_k}}}{{{b_k}}}\frac{1}{{\left( {{\pi ^2} + d_k^2h_{\,m}^{(k)2}} \right)}},\,\,\,m = 1,} \\ 
  {\frac{{{d_k}}}{{{b_k}{d_k}h_m^{(k)}\sinh({d_k}h_m^{(k)})}},\,\,\,\,\,\,\,\,\,\,\,\,\,\,\,\,\,\,\,\,\,\,\,\,\,\,\,\,\,\,\,\,\,\,\,\,\,\,\,\,\,\,\,\,\,m \ne 1.} 
\end{array}} \right.\]               

\[R_s^{(k,1)} = \frac{1}{2}\frac{{{J_0}\left( {\frac{{{a_k}{\lambda _1}}}{{{b_k}}}} \right)}}{{{{\left( {\frac{{{a_k}{\lambda _1}}}{{{b_k}}}} \right)}^2} - {{\left( {{\lambda _s}} \right)}^2}}},\,\,\,\,R_s^{(k,2)} = \frac{1}{2}\frac{{{J_0}\left( {\frac{{{a_k}{\lambda _1}}}{{{b_{k - 1}}}}} \right)}}{{{{\left( {\frac{{{a_k}{\lambda _1}}}{{{b_{k - 1}}}}} \right)}^2} - {{\left( {{\lambda _s}} \right)}^2}}},\]

\[W_{s',s}^{(1)} = \frac{{a_1^2}}{{b_{w,1}^2}}\sum\limits_{s''} {\,\frac{{{\lambda _{s''}}}}{{J_1^2\left( {{\lambda _{s''}}} \right)\gamma _{s''}^{(1)}{b_{w,1}}}}\frac{{{J_0}\left( {\frac{{{a_1}{\lambda _{s''}}}}{{{b_{w,1}}}}} \right){j_{2s - 1}}\left( {\frac{{{\lambda _{s''}}{a_1}}}{{{b_{w,1}}}}} \right)}}{{{{\left( {\frac{{{a_1}{\lambda _{ss''}}}}{{{b_{w,1}}}}} \right)}^2} - \lambda _{s'}^2}}} ,\,\,\,\,\,{V_{s'}} = \frac{{{J_0}\left( {\frac{{{a_1}{\lambda _1}}}{{{b_{w,1}}}}} \right)}}{{{{\left( {\frac{{{a_1}{\lambda _1}}}{{{b_{w,1}}}}} \right)}^2} - \lambda _{s'}^2}}\]

\[W_{s',s}^{(2)} = \frac{{a_{{N_R} + 1}^2}}{{b_{w,2}^2}}\sum\limits_{s''} {\,\frac{{{\lambda _{s''}}}}{{J_1^2\left( {{\lambda _{s''}}} \right)\gamma _{s''}^{(2)}{b_{w,2}}}}\frac{{{J_0}\left( {\frac{{{a_{{N_R} + 1}}{\lambda _{s''}}}}{{{b_{w,2}}}}} \right){j_{2s - 1}}\left( {\frac{{{\lambda _{s''}}{a_{{N_R} + 1}}}}{{{b_{w,2}}}}} \right)}}{{{{\left( {\frac{{{a_{{N_R} + 1}}{\lambda _{s''}}}}{{{b_{w,2}}}}} \right)}^2} - \lambda _{s'}^2}}} \]

\end{document}